\documentclass[aps,pre,]{revtex4-2}

\usepackage[utf8]{inputenc}
\usepackage{amsmath}
\usepackage{amssymb}
\usepackage{graphicx}
\usepackage{siunitx} 
\usepackage{tikz}
\usepackage{amsfonts}
\usepackage{subcaption}
\usepackage{xr}
\makeatletter
\newcommand*{\addFileDependency}[1]{
  \typeout{(#1)}
  \@addtofilelist{#1}
  \IfFileExists{#1}{}{\typeout{No file #1.}}
}
\makeatother

\newcommand*{\myexternaldocument}[1]{%
    \externaldocument{#1}%
    \addFileDependency{#1.tex}%
    \addFileDependency{#1.aux}%
}
\myexternaldocument{supplementary/suppMat}
\newcommand{\clusterA}{ \tikz[baseline=10pt,scale=0.8]{
\draw (0,0) node[circle,draw,fill,inner sep=0pt,minimum size=4pt] {} -- (0,1) node[circle,draw,fill,inner sep=0pt,minimum size=4pt] {};
}}
\newcommand{\clusterB}{ \tikz[baseline=10pt,scale=0.8]{
\draw (0,0) node[circle,draw,fill,inner sep=0pt,minimum size=4pt] {} -- (0,1) node[circle,draw,fill,inner sep=0pt,minimum size=4pt] {};
\draw (0,1) node[circle,draw,fill,inner sep=0pt,minimum size=4pt] {} -- (1,1) node[circle,draw,fill,inner sep=0pt,minimum size=4pt] {};
}}
\newcommand{\clusterC}{ \tikz[baseline=10pt,scale=0.8]{
\draw (0,0) node[circle,draw,fill,inner sep=0pt,minimum size=4pt] {} -- (0,1) node[circle,draw,fill,inner sep=0pt,minimum size=4pt] {};
\draw (0,1) node[circle,draw,fill,inner sep=0pt,minimum size=4pt] {} -- (1,1) node[circle,draw,fill,inner sep=0pt,minimum size=4pt] {};
\draw (1,1) node[circle,draw,fill,inner sep=0pt,minimum size=4pt] {} -- (1,0) node[circle,draw,fill,inner sep=0pt,minimum size=4pt] {};
}}
\newcommand{\clusterD}{ \tikz[baseline=10pt,scale=0.8]{
\draw (-0.5,0,0) node[circle,draw,fill,inner sep=0pt,minimum size=4pt] {} -- (0,1) node[circle,draw,fill,inner sep=0pt,minimum size=4pt] {};
\draw (0,1) node[circle,draw,fill,inner sep=0pt,minimum size=4pt] {} -- (0.5,0) node[circle,draw,fill,inner sep=0pt,minimum size=4pt] {};
\draw (0.5,0) node[circle,draw,fill,inner sep=0pt,minimum size=4pt] {} -- (-0.5,0) node[circle,draw,fill,inner sep=0pt,minimum size=4pt] {};
}}
\DeclareMathOperator{\tr}{Tr\,}

\begin{document}

\title{The native state of natural proteins optimises local entropy}

\author{M. Negri}
\affiliation{Department Applied Science and Technology, Politecnico di Torino, CorsoDuca degli Abruzzi 24, I-10129 Torino, Italy}
\email{matteo.negri@polito.it}
\author{G. Tiana}
\affiliation{Department of Physics and Center for Complexity and Biosystems, Universit\`a degli Studi di Milano and INFN, via Celoria 16, 20133 Milano, Italy}
\email{guido.tiana@unimi.it}
\author{R. Zecchina}
\affiliation{Artificial Intelligence Lab, Bocconi University, Via Sarfatti, 25, 20136 Milano, Italy}
\email{riccardo.zecchina@unibocconi.it}
\date{\today}

\begin{abstract}
The differing ability of polypeptide conformations to act as the native state of proteins has long been rationalized in terms of differing kinetic accessibility or thermodynamic stability. Building on the successful applications of physical concepts and sampling algorithms recently introduced in the study of disordered systems, in particular artificial neural networks, we quantitatively explore how well a quantity known as the local entropy describes the native state of model proteins. In lattice models and all-atom representations of proteins, we are able to efficiently sample high local entropy states and to provide a proof of concept of  enhanced stability and folding rate. Our methods are based on simple and general statistical--mechanics arguments, and thus we expect that they are of very general use.
\end{abstract}

\maketitle

\section{Introduction}
Proteins are the machinery of life. In order to perform their functions, the majority of proteins fold into a compact native state that is intimately linked with their polypeptide conformation, as it has been known for thirty years \cite{Finkelstein1993}.
The goal of this study was to rationalize the observation that the number of protein sequences is largely more abundant than the number of protein conformations, causing substantial degeneracy in the map between sequence and structure. The problem has regained importance in more recent years, when it has become feasible to design proteins {\it de novo} with custom functions, and thus it has become critical to understand which conformations can be best designed \cite{Chevalier2017}.

Most theoretical results point to the conclusion that we can observe a small subset of all possible protein conformations because they exhibit physical properties that make them biologically more fit rather than just because they are poorly sampled by evolution. Some conformations therefore appear to be more "designable" than others \cite{Li1996}.
Some works justify the better designability of existing conformations by their greater kinetic accessibility during the folding process, often associated with conformational symmetries such as secondary structures \cite{Govindarajan1996,Wolynes1996,Maritan2000}. An optimal balance between local and nonlocal contacts would make the folding rate of some native conformations particularly fast. This hypothesis is supported by the correlation observed in proteins between the average linear separation between residues in contact in the native state and the rate of folding \cite{Plaxco1998}.
Other works relate the design of a native conformation to its thermodynamic stability, thus focusing on its equilibrium rather than kinetic characteristics. The basic idea is that stable proteins exhibit a large gap between the energy of the native state and those of competing conformations \cite{Shakhnovich1994} and this gap can accommodate a large number of sequences that fold to the same native state \cite{Broglia1999,Tiana2001b}. Different conformational properties may contribute to this enhanced thermodynamic stability. More compact conformations are more stable because they can exhibit more attractive interactions and because they protect more efficiently hydrophobic residues from the solvent. Daisy-like conformations that maximize the trace of the eighth power of the contact matrix have also been shown to be particularly stable \cite{England2003}; in fact, this quantity has been shown to correlate with the evolutionary age of the proteins \cite{Tiana2004d}. Similarly, the presence of loops of specific sizes has been found to improve thermodynamic stability and justified by \cite{Berezovsky2000} energy arguments.

Still, the fact that specific protein conformations can be particularly stable at equilibrium is usually associated with the property of amino acid sequences to exhibit markedly low potential energy in those conformations, thus more efficiently minimizing system frustration.  On the other hand, although the native state of proteins is usually considered macroscopically unique, its entropy is not negligible compared to the competing denatured state \cite{Karplus1987}. This entropy arises from the constellation of conformations, structurally similar to the ground state, that lie beyond the transition state. They certainly include several vibrational states, conformational sub-states \cite{Frauenfelder1988}, and perhaps other conformations whose contribution to the partition function cannot be separated from that of the ground state.

The hypothesis we wish to further investigate in the present work is whether the entropy of the native state of natural proteins, and not just the potential energy, is particularly optimized compared to that of random conformations. This hypothesis was suggested several years ago~\cite{Shortle1998a} by a qualitative computational analysis in which some of the most mobile dihedrals of small proteins were changed and the number of conformations within 4\AA  ~from native conformations and devoid of steric clashes was estimated. It was found that natural proteins exhibit more neighboring conformations than random decoys. More recently, a knowledge--based local--entropy parametrization was used to predict contact changes between amino acids during protein conformational changes \cite{Sankar2017}.

From a statistical--physics point of view, this quantity is captured by the so-called {\em local entropy}, i.e., the log of the number of low energy configurations within a given distance from a reference configuration (for continuous systems the definition can be generalized straightforwardly).
Recently, it has been seen that the notion of local entropy plays a central role in systems that exhibit a potentially very complex energy landscape and at the same time possess highly accessible states that correspond to high local entropy minima \cite{baldassi2015subdominant,unreasoanable}.  The latter turn out to be accessible by a multitude of dynamical processes which are not designed to have the Gibbs distribution as stationary probability measure, due e.g. to non-thermal external perturbations. Systems of this type are non-convex models of artificial neural networks (including deep neural networks), in which entropic phenomena play essential roles \cite{baldassi2020shaping} for the (unexpected) success of the current learning processes, largely based on non-equilibrium variants of gradient descent.

Here we  want to generalize the algorithmic schemes introduced for  sampling high local entropy ground states in neural systems \cite{baldassi_local_2016} and apply them to simple models of 3D protein structures. Considering both lattice model and all-atom representations of proteins, we show that by sampling native states with high local entropy (which are in principle rare compared to states that dominate the Gibbs measure) we find a decrease in the linear separation between contact residues.  In addition, the "flatness" of the energy profile in the native state can extend to the transition state, having consequences both on the thermodynamic stability of the protein, lowering the free energy of the native state, and on the folding rate, lowering the free energy of the transition state.
The generality of the sampling method would allow it to be used in conjunction with any structure prediction method, such as e.g. Alpha--fold \cite{jumper2021highly}   and Rosetta Fold \cite{baek2021accurate}, and could aid in the search for sequences folding onto the most designable structures.

\section{Local entropy and how to calculate it}
\label{sect:locent}

To define a probability measure that ignores narrow ground states and enhances the statistical weight of large dense regions of ground states, we can consider the  local free-entropy
\begin{equation}
    S_{\mathrm{loc}}(\Gamma,\gamma,\beta)=\log \int d\Gamma' \exp\left[-\beta U(\Gamma')- \gamma d(\Gamma,\Gamma')\right],
    \label{eq:locent}
\end{equation}
where $d (\Gamma,\Gamma')$ is any metrics suitable for the model under consideration and $\gamma$ is its conjugate Lagrange multiplier. We describe the explicit choice of $d$  in section \ref{sect:homo} (see Eq.~\ref{eq:dist}). Here and in the following, we shall set Boltzmann's constant to 1.
In the limit of $\beta \to \infty$, this expression reduces (up to an additive constant) to a "local entropy": it counts the number of minima $\Gamma'$ of the energy, weighing them (via the parameter $\gamma$) by the distance to a reference configuration $\Gamma$. For continuous variables, the local entropy becomes the log of a weighted volume around a reference configuration. We can then define the probability distribution
\begin{equation}
P(\Gamma;y,\gamma, \beta)=\frac{1}{Z(y,\gamma,\beta)} e^{y S_{\mathrm{loc}}(\Gamma,\gamma,\beta)}
    \label{eq:P_locent}
\end{equation}
where $y$  determines the degree of concentration of the probability distribution on high local entropy regions.
When $y$ is large, only the configurations $\Gamma$  that are surrounded by an large number of local minima will have non-negligible weight. By increasing the value of $\gamma$, it is possible to focus on tighter neighborhoods around $\Gamma$, and at large values of $\gamma$ the target $\Gamma$ will also share with high probability the properties of the surrounding minima.
From an algorithmic perspective we can use the high local entropy probability distribution as a starting point for designing a Markov Chain, in the same way that simulated annealing uses the Gibbs measure. One possibility \cite{unreasoanable,baldassi_local_2016} is to observe that if we take $y$ to be a non-negative integer we can rewrite the partition function as a product of identical systems connected by a distance constraint
\begin{equation}
\begin{split}
&Z (y,\gamma,\beta)=\int d\Gamma e^{y S_{\mathrm{loc}}(\Gamma,\gamma,\beta)}  \\
&=\int d\Gamma_c  \prod_{a'=1}^y d\Gamma_{a'} \;  e^{-\beta \sum_{a=1}^y  U(\Gamma_a)+ \gamma \sum_{a=1}^y d(\Gamma_a,\Gamma_c) -\beta  U(\Gamma_c)}
    \label{eq:Z_locent}
\end{split}
\end{equation}
where $\Gamma_c$ is the "central" reference configuration, and $\{\Gamma_a\}$ are the configurations of the replicated systems.
This partition function describes a system of $y + 1$ interacting replicas of the initial system, one of which acts as the reference system, while the other $y$ systems are subject to a distance constraint with respect to the reference system. This gives us a very simple and general scheme to direct algorithms to sample wide minima  of the  the energy landscape: replicate the model, add an interaction term with a reference configuration, and run the algorithm on the resulting extended system. In practice we only need to consider the following effective system
\begin{equation}
    U_{\text{eff}}(\Gamma_c,\{\Gamma_a\})=\sum_{a=1}^y \left[ U(\Gamma_a)- \frac{\gamma}{\beta} d(\Gamma_a,\Gamma_c) \right]+U(\Gamma_c)
    \label{eq:center}
\end{equation}
and run our preferred Monte Carlo (MC) Markov Chain update. Replicas are initialized at random, while the center configuration is initialized at the average of the replica configurations. It is worth noting that $y$ controls the value of the local entropy and that relatively small values of $y$ are sufficient to obtain the results we are interested in. By taking the inverse temperature  of the individual systems $\beta$ to be large, we  focus the sampling on flat ground states.

\section{Effects of the local entropy on the equilibrium conformations of polymers}
\label{sect:homo}

We first tested the effect of controlling the local entropy of a polymer on a simple cubic--lattice model, which offers the advantage of a fast sampling of the conformational space of the system and of defining unambiguously the zero of the entropy. We employed a standard model on a cubic lattice \cite{Shakhnovich1994} in which the beads, sitting on the vertices of the lattice, cannot overlap. They interact with a contact potential depending on the conformation $\Gamma=\{\vec{x}_i\}_{i=1}^N$
\begin{equation}
    U(\Gamma)=\frac{1}{2}\sum_{ij=1}^N J_{ij} \Delta_{ij}(\Gamma),
    \label{eq:ulattice}
\end{equation}
where the contact function $\Delta_{ij}(\Gamma)$ is 1 if the $i$th and $j$th beads are neighbours in space and $|i-j|>2$ and zero otherwise.

The chain is simulated with a standard Monte Carlo algorithm that includes the corner flip, the crankshaft and rotations of the ends as elementary moves. At each step, a monomer is chosen at random for it with flat {\it a priori} probability and a random move is chosen randomly with uniform probability among those that are possible, accepting it with the standard Metropolis probability. The initial conformation is generated from a random self--avoiding walk in the lattice. The simulations for computing equilibrium quantities consist of $10^7$ steps, recording the conformation every $10^4$ steps. See Fig.~S1 \cite{suppMat} for example trajectories that reach equilibrium. Simulations we use to compute the folding time consist of $10^6$ steps, recording the conformation every $10^2$ steps.

In order to control the local entropy of the system, we performed a Monte Carlo simulations of $y$ replicas of the system starting from independent conformations and interacting with the potential defined in Eq. (\ref{eq:center}). The Metropolis acceptance rate now reads $p_{\mathrm{accept}}=\min \left( 1, e^{-\Delta} \right)$, where $ \Delta = \beta \Delta E + \gamma \Delta d$ and
\begin{equation}
    \Delta d =  \begin{cases}
                    d(\Gamma_a^\mathrm{new},\Gamma_c) - d(\Gamma_a^\mathrm{old},\Gamma_c), & \text{if we are moving a replica $a$}\\
                    \frac{1}{y}\sum_y [d(\Gamma_a,\Gamma_c^\mathrm{new}) - d(\Gamma_a,\Gamma_c^\mathrm{old})], & \text{if we are moving the center $c$}.
                \end{cases}
\end{equation}

\subsection{Increased local entropy depletes long--range contacts in homopolymers}

The simplest polymer model that can be studied is the lattice homopolymer, obtained setting $J_{ij}=-1$ in Eq. (\ref{eq:ulattice}) for each pair $i,j$. We compared the results of a standard Monte Carlo sampling of the Boltzmann distribution with a replicated Monte Carlo, which controls the local entropy through the parameter $\gamma$, as described in Sect. \ref{sect:locent}. To define the local free entropy we adopt for this system a distance function defined by
\begin{equation}
    d(\Gamma_1,\Gamma_2)=1-\frac{1}{N_c}\sum_{i<j}^N\Delta_{ij}(\Gamma_1)\Delta_{ij}(\Gamma_2),
\label{eq:dist}
\end{equation}
that is the fraction of different contacts between the two conformations $\Gamma_1$ and $\Gamma_2$; here $N_c$ is the maximum number of contacts that the chain can build.
As displayed in Fig.~S2 in Supp. Mat., we have checked that the results are robust with respect to the distance function (e.g. by comparison with root mean square distance, RMDS) and to a different coupling scheme of the replicas.
What might be an optimal definition of distance is an interesting problem that goes beyond the scope of our  study and that might deserve further study.

\begin{figure*}
    \centering
    \subcaptionbox{\hspace*{6cm}}{\includegraphics[width=0.49\linewidth]{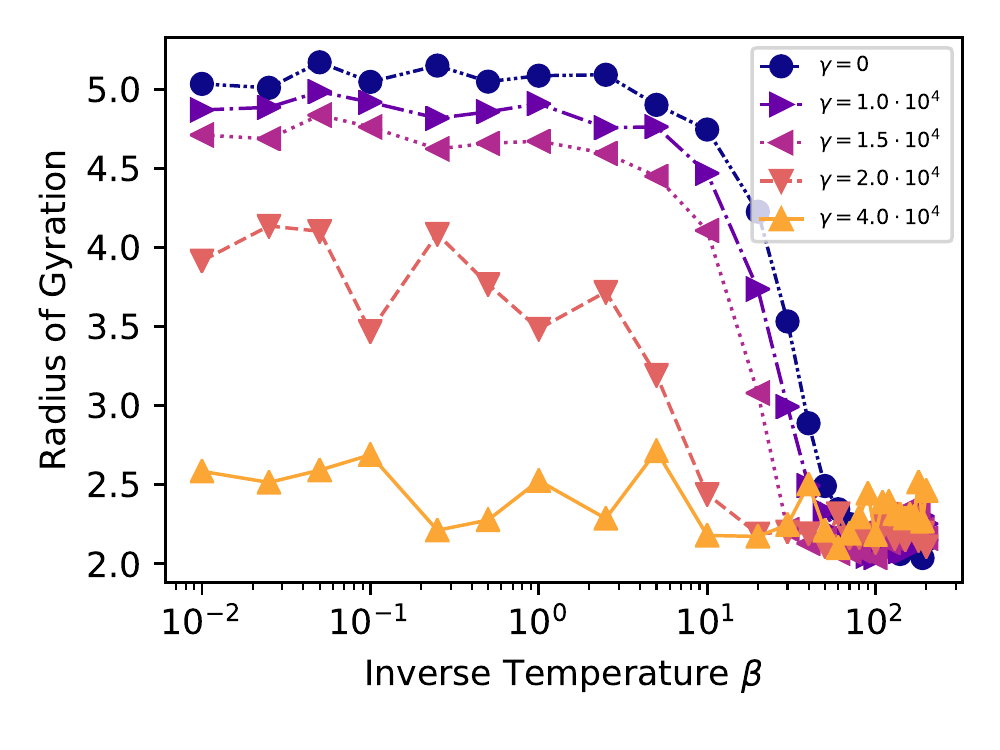}}
    \subcaptionbox{\hspace*{6cm}}{\includegraphics[width=0.49\linewidth]{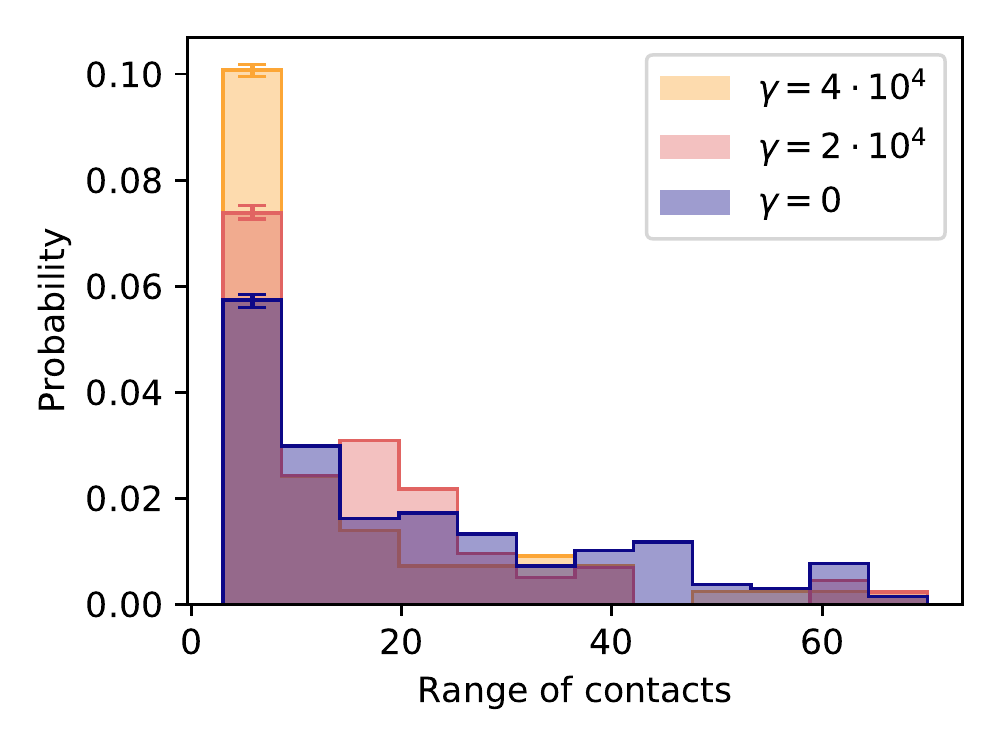}}

    \caption{\textbf{Local entropy changes the properties of a lattice homopolymer.} (a) Radius of gyration as a function of the inverse temperature for increasing values of $\gamma$. $R_g$ is dimensionless, since measure lengths in units of the lattice step which is set to 1. The points represent thermal averages. We used $y=3+1$ replicas, the last one being the center.  (b) Distribution of the range of contacts $|i-j|$ at $\beta=100$ for increasing values of $\gamma$. Increasing $\gamma$ suppresses long-range contacts and favours short-range ones. The error bars on the fist bin are generated by bootstrapping.}
    \label{fig:homopol}
\end{figure*}

In a homopolymer the only relevant equilibrium effect is the coil--globule transition. A comparison between the transition described by a Boltzmann sampling and that obtained controlling the local entropy (cf. Sect. \ref{sect:locent}) is shown in Fig.~\ref{fig:homopol}a for a polymer with $N=70$ (see Fig.~S3 in Supp. Mat. for two example configurations). In the case of Boltzmann sampling ($\gamma=0$) the coil--globule transition is marked by a jump of the radius of gyration ($R_g$) at an inverse temperature $\beta\approx 30$. In the lattice model we measure lengths in units of the lattice step, which is set to 1 so that $R_g$ is dimensionless.
Increasing the bias associated with the local entropy (i.e., by increasing $\gamma$) leads to a stabilization of the globule, the transition temperature increasing with $\gamma$. At the same time, we observe a compaction of the coil state, associated with the fact that at short distances  more compact conformations have certainly a larger number of neighbouring other conformations (cf. Eq. \ref{eq:dist}) and thus a larger local entropy. The globular phase, in which the polymer is maximally compact, is weakly affected.

In  Fig.~\ref{fig:homopol}b we compared the distributions of contact range (i.e., of the values of $|i-j|$ when $i$ and $j$ are in contact) for fixed inverse temperature $\beta=100$, at which the system is in the globule phase at all values of $\gamma$ we simulated. A homopolymeric globule is expected to display an initial decrease of the contact probability up to the distance corresponding to the diameter of the globule, followed by a flat distribution \cite{DeGennes1979a}.  At $\gamma=0$ the distribution displays overall the expected character but is quite irregular because of the constraints imposed by the cubic lattice on the $R_g$ of highly--compact conformations (i.e., there is a non--monotonic relation between $R_g$ and energy).  Increasing $\gamma$, we observed a regular increase of short--range contacts at the expense of long--range contacts.

\subsection{Increased local entropy simultaneously stabilizes and decreases folding time of model proteins}
\label{subsect:time}

To obtain a clear picture of the effects of the local entropy on the folding properties of model proteins, we calculated the equilibrium stability and the folding rate of a Go model \cite{Go1983} on lattice. This is defined choosing a target conformation $\Gamma_0$ and setting
\begin{equation}
J_{ij}=-J\cdot\Delta_{ij}(\Gamma_0)
\label{eq:j}
\end{equation}
in Eq. (\ref{eq:ulattice}), where $J$ defines the energy scale and was set to 1. With this choice of $J_{ij}$, the target conformation is by definition the ground state of the system, and thus the equilibrium state at low temperature. The reason for using a Go model is to decouple the effect of protein sequence from that of protein structure, focusing our attention only on the latter. Here the protein sequence is described effectively, assuming that evolution has minimized energetic frustration to the maximum degree \cite{Bryngelson1987}.

We chose as target conformation $\Gamma_0$ either conformation sampled by the homopolymeric model according to Boltzmann distribution ($\beta=100$ and $\gamma=0$) or biasing their local entropy ($\beta=100$ and $\gamma>0$), in all cases taking care of selecting only globular conformations (our choice for the cutoff is $R_g < 2.5$). We thus obtained different potentials that depend on $\gamma$ through the choice of $\Gamma_0$.

We then simulated at low temperature ($\beta=120$) the dynamics of the system starting from a random, high--temperature, coil state with the standard Metropolis scheme, that at fixed temperature approximates the Smoluchowski equation and thus reports realistic trajectories of the system \cite{Tiana2007b}. From each trajectory we obtained the fraction of native contacts $f_N(t)$ (that for the Go model is $=U(\Gamma(t))/U(\Gamma_0)$, where $\Gamma(t)$ is the conformation of the chain at time $t$), we calculated the average $\overline{f_N(t)}$ over 40 simulations and fitted them by a two--state kinetics
\begin{equation}
    f_N(t)= f_{\mathrm{eq}} \cdot (1-e^{-t/\tau}) +f_0
\end{equation}
where $\tau$ is the mean folding time, $f_0$ is the residual fraction of native contacts in the initial conformation,  $f_{eq}$ is the equilibrium similarity to the target conformation, that in the two--state approximation is equal to the equilibrium probability of the native state (cf. Fig.~S4 in the Supp. Mat.). We show in Table~S1 in Supp. Mat. the average and standard deviation of the mean square error for each fitted curve.

\begin{figure*}
    \subcaptionbox{\hspace*{6cm}}{\includegraphics[width=0.485\linewidth]{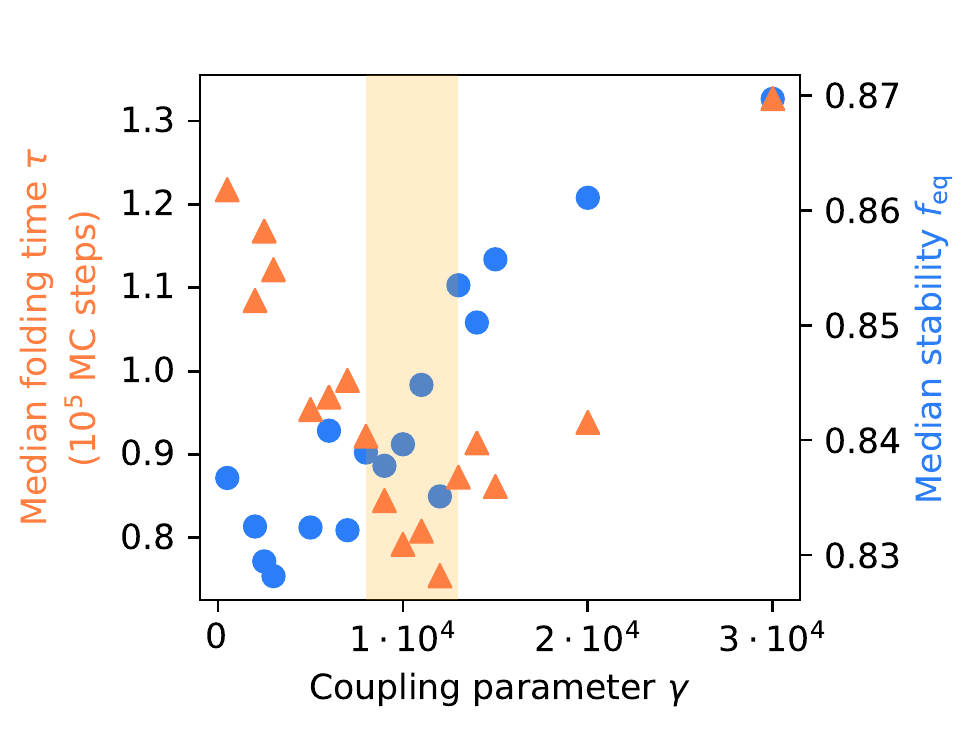}}
    \subcaptionbox{\hspace*{8cm}}{\includegraphics[width=0.505\linewidth]{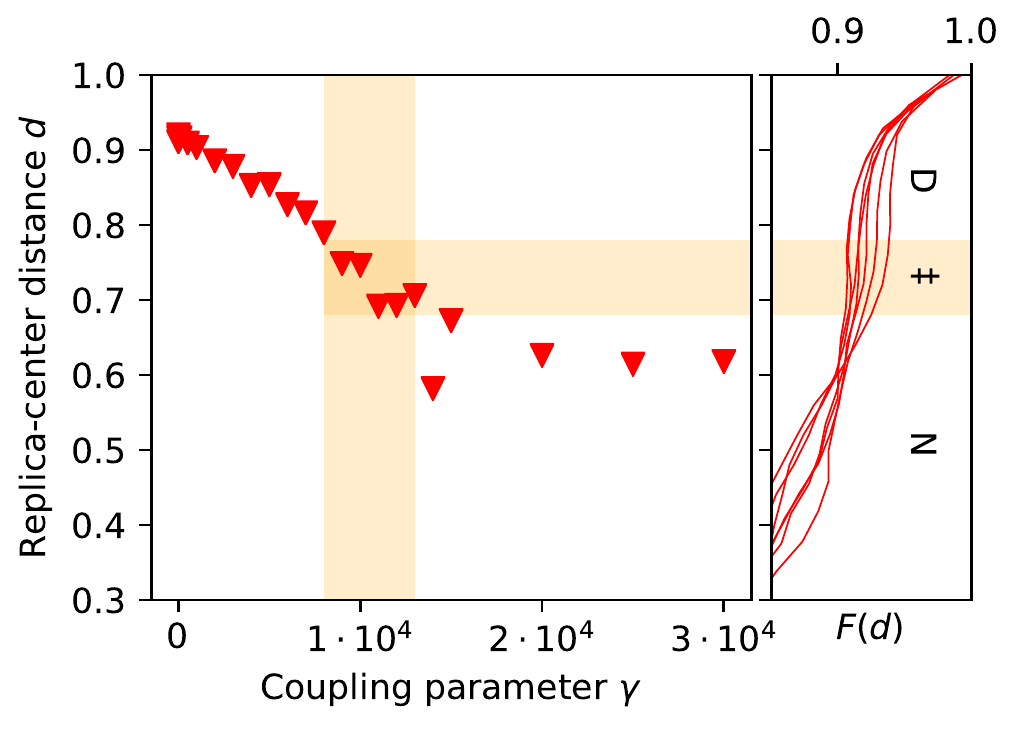}}
    \caption{\textbf{Structures with high local entropy are more stable and fold faster.} (a) The average folding time (in MC steps, orange triangles, left y-axis) for target conformations $\Gamma_0$ obtained at different values of $\gamma$ and the average stability of the structure (blue circles, right axis). The orange bar indicates the transition state $\ddagger$ calculated in the right panel. (b) The average distance between the central conformation and the replicas at different values of $\gamma$ obtained from the simulation used to generate the $\Gamma_0$. Aligned with the right axis we show the free energy of a Go model as function of the distance from the native state for various structures at low $\gamma$. Both the distance and the free energy are dimensionless given the definition of the model. We can see that the region around $\gamma=1\cdot 10^4$, marked with an orange bar, corresponds to the typical distance between native and transition state.}
    \label{fig:folding_time}
\end{figure*}

In  Fig. ~\ref{fig:folding_time}a we plotted the values of $\tau$ (in MC steps) and $f_{\mathrm{eq}} $ as a function of the parameter $\gamma$ that controls the local entropy of the target conformation $\Gamma_0$. The values are medians over 60 realization of $\Gamma_0$. The corresponding standard deviations are shown in the Supp. Mat., Table~S1. It is shown that the stability of the native state increases with $\gamma$, while the folding time displays a non--monotonic behaviour with a minimum at $\gamma\approx 1.2\cdot 10^4$.

To rationalize these results, we estimated what is the radius of the neighborhood of the target conformation that is affected by the increase in local entropy at varying $\gamma$. In  Fig.~\ref{fig:folding_time}b we plotted the average inter--conformation distance obtained from the replica simulation from which we obtained the conformations $\Gamma_0$ (cf. Eq.~\ref{eq:center}) as a function of $\gamma$. For large $\gamma$ ($>1.5\cdot 10^4$) the average distance displays a plateau at $d\approx 0.6$; decreasing $\gamma$ the average distance increases, overcoming 0.9 for plain Monte Carlo simulations. In the plot is also marked the transition state at $d=0.7$, that separates the denatured state from the native basin.

Both the stabilization and the kinetic effects of the local entropy can be clarified noting what part of the free--energy profile $F(d)=E(d)-TS(d)$ is affected by the local entropy. At large $\gamma$ the local entropy affects only the close neighborhood of the $\Gamma_0$ ($d\lesssim0.4$), decreasing its free energy and thus stabilizing it with respect to the denatured state, which is unaffected. The transition state is not affected as well, so the folding rate, that according to Kramers theory depends on the free energy of the transition state calculated with respect to the denatured state, is similar to that at $\gamma=0$. When $\gamma$ is decreased, the neighbourhood of $\Gamma_0$ affected by the increase in local entropy reaches the transition state ($d\simeq0.7$) and lowers it, decreasing the folding time. The non--monotonicity of the folding time arises from the fact that making $\gamma$ even smaller, also the denatured state is affected and then folding barrier grows again.

The degree of cooperativity of the folding transition decreases slightly with $\gamma$ (see Table~S2 in Supp. Mat.). This is not unexpected because $\gamma>0$ leads to the stabilization of conformations with varying distance from the native one, thus decreasing the two-state character of the folding transition.

\section{The local entropy of natural proteins is larger than that of random decoys}
\label{sec:md}

We then tested the hypothesis that the native state of natural proteins displays a larger local entropy than random polypeptidic conformations with the same density. We studied seven natural proteins and a stable protein designed {\it de novo} (HHH) \cite{Chevalier2017}.

To estimate the local entropy associated with the native state we need to describe the energy of the protein in the neighborhood of the crystallographic structure. For this purpose we made use of an all--atom Go model \cite{Go1983}, that is expected to be particularly realistic in the native basin. At the same time, it allows one to decouple the effect of the sequence from that of the native conformation in the calculation of the local entropy.

In order to evaluate the local entropy, we performed molecular dynamics (MD) simulations with Gromacs 2020.4 \cite{abraham2015gromacs} using the all--atom Go model obtained by Smog2 \cite{noel2016smog}. The native conformations of the proteins were: protein G (pdb code 1pgb, 56 residues), ACBP (pdb code 2abd, 87 residues), CI2 (pdb code 2ci2, 83 residues), src-SH3 (pdb code 1srl, 64 residues), villin headpiece (pdb code 5vnt, 63 residues), barnase (pdb code 1bnr, 110 residues) and HHH (artificial protein, pdb code 5uoi, 43 residues).
MD simulations were performed in the range of temperature from 1 to 200 (in energy units) for $2\cdot10^5$ steps of time step $5\cdot 10^{-4}$ ps with stochastic dynamics \cite{abraham2015gromacs}.
The Go potential of each decoy, whose parameters in Smog2 depend on the number of residues and of native contacts in the native conformation, is rescaled to that of the corresponding native protein in order to facilitate the comparison among them.
The microcanonic entropy $S(E)$ is extracted from all the simulations performed at different temperatures for the same protein  with the maximum--likelihood code developed in ref. \cite{tiana2011equilibrium}.

As a control model we generated putative native conformations from a homopolymeric model, derived from the original models as follows. Starting from the crystallographic structure of each protein, we generated an all-atom model similar to that described above (i.e., with the same atomic structure), but where each pair of atoms interacts instead in the same way with the Lennard--Jones potential
\begin{equation}
    V_\mathrm{LJ}(r_{ij})= C^{(12)}_{ij}/r_{ij}^{12} -C^{(6)}_{ij}/r_{ij}^{6}
\end{equation}
where for all $i,j$ we set $C^{(6)}_{ij}=1.4\cdot10^{-2} \,\text{kJ}\,\text{mol}^{-1}\,\text{nm}^6$ and $C^{(12)}_{ij}=1.0\cdot10^{-4} \,\text{kJ}\,\text{mol}^{-1}\,\text{nm}^6$. No potential is applied to the dihedrals at this stage. The parameters $C^{(6)}$ and $C^{(12)}$ are chosen with a grid search so that, after an annealing MD of the chain, the number of contacts in the putative conformations is not smaller than in the original model (see Fig.~S5 in the Supp. Mat.), to rule out trivial effects in the calculation of the local entropy. The resulting conformation is used as putative $\Gamma_0$ (examples of these configurations can be found in Fig.~S6 in the Supp. Mat.).

\begin{figure*}
    \centering
    \subcaptionbox{\hspace*{6cm}}{\includegraphics[width=0.4\linewidth]{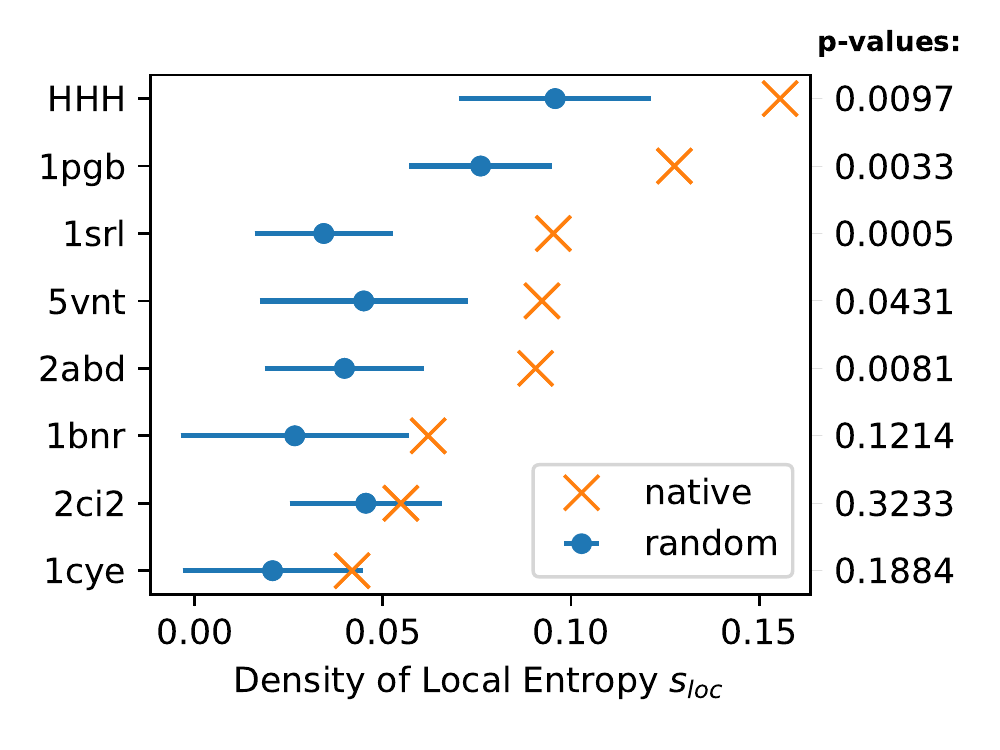}}
    \subcaptionbox{\hspace*{6cm}}{\includegraphics[width=0.4\linewidth]{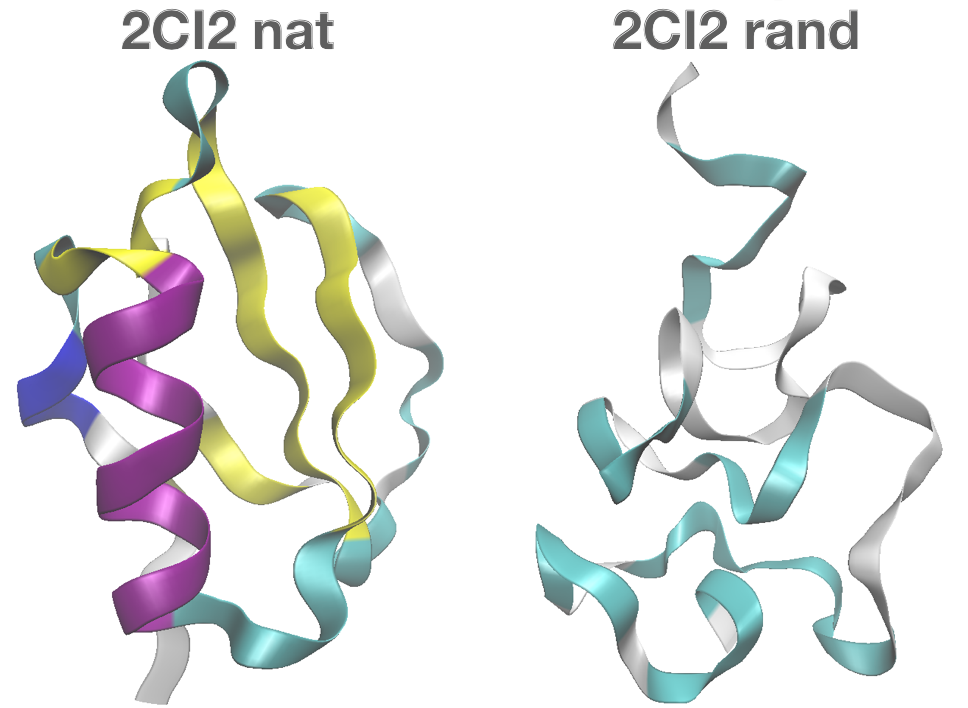}}
    \subcaptionbox{\hspace*{6cm}}{\includegraphics[width=0.4\linewidth]{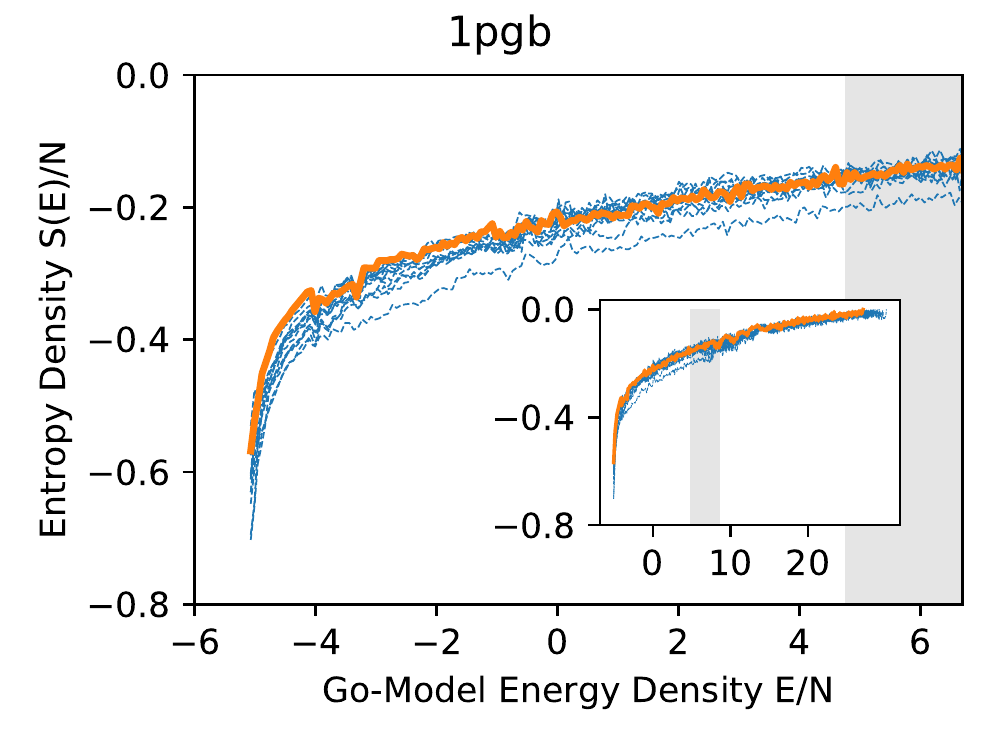}}
    \subcaptionbox{\hspace*{6cm}}{\includegraphics[width=0.4\linewidth]{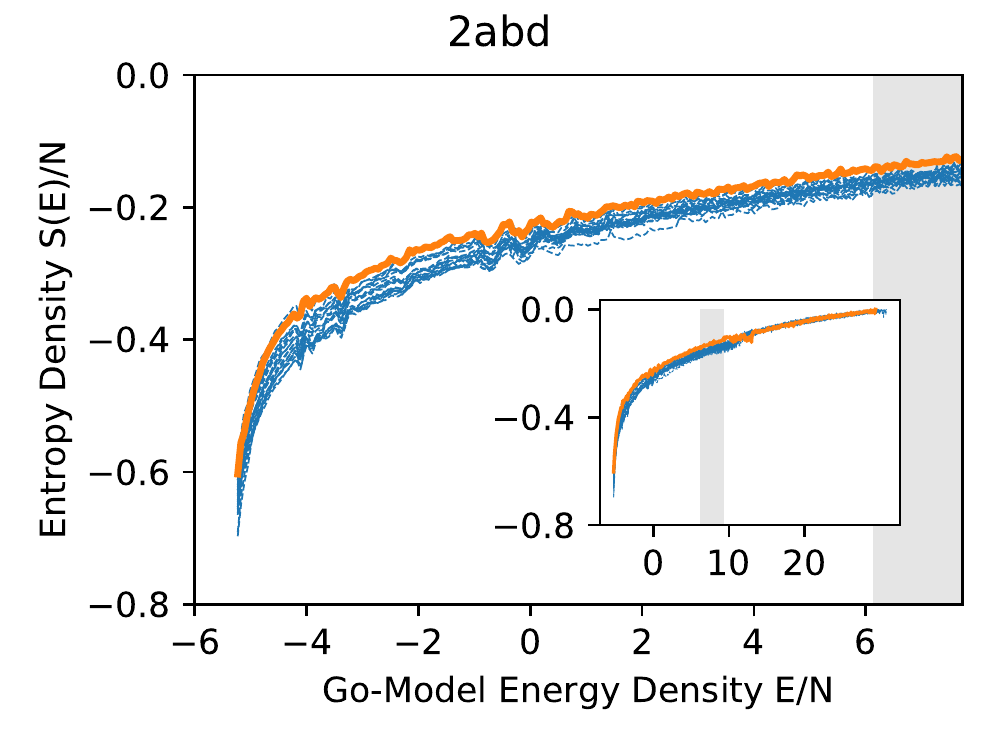}}
    \subcaptionbox{\hspace*{6cm}}{\includegraphics[width=0.4\linewidth]{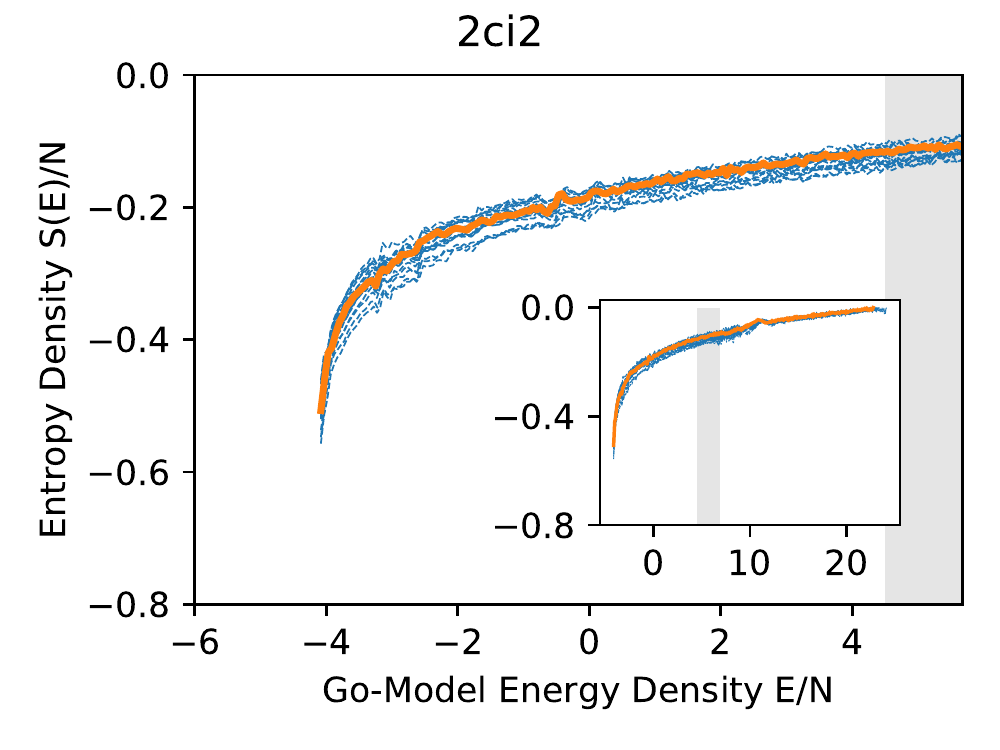}}
    \subcaptionbox{\hspace*{6cm}}{\includegraphics[width=0.4\linewidth]{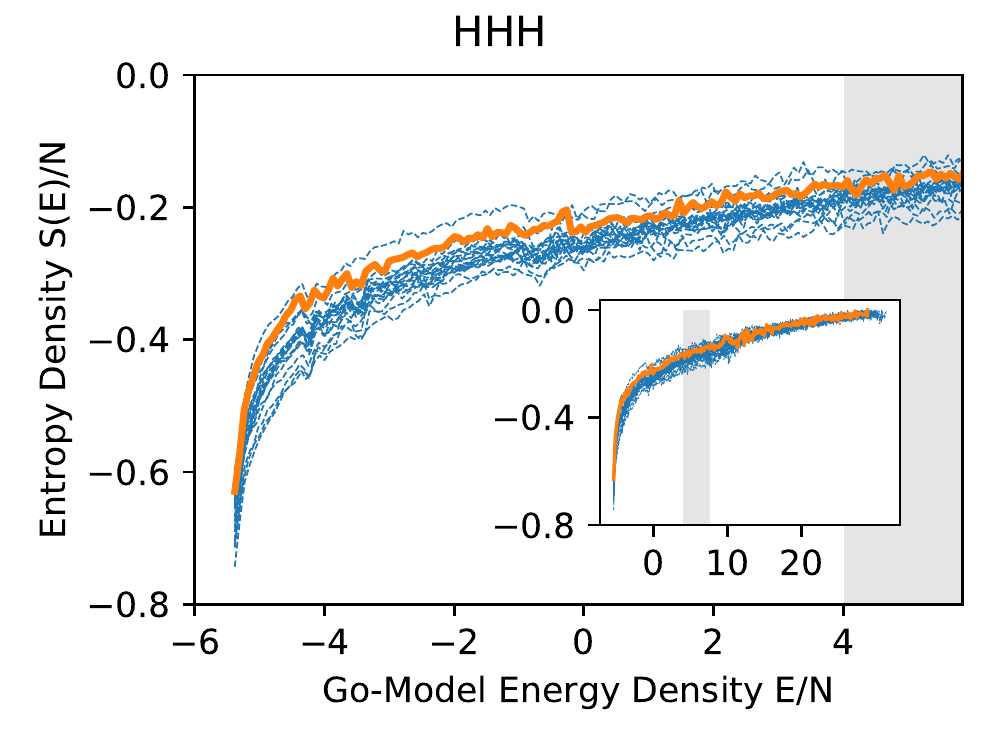}}
    \caption{\textbf{Natural structures have higher local entropy than random ones at low energy.} (a) The density of local entropy of the native state (orange crosses), the average density of local entropies of the decoys (blue circles) and the associated standard deviation (blue bars) for each protein. On the right axis are reported the associated p--values. (b) the native conformation of CI2 (2ci2) and an example of decoy. (c--e) The entropy density calculated as a function of energy for four proteins (solid orange curves) and for the associated decoys (dashed blue curves). The grey vertical bars indicate the region where the transition states are (the band is centered on the average over the protein and the decoys, its width is twice a standard deviation). In the inset is displayed the whole entropy density, while in the main figures only the region of the native state. Note the energy density has been rescaled for every structure as described in section \ref{sec:md}, so the energy units are arbitrary.}
    \label{fig:flatness}
\end{figure*}

The entropy $S(E)$ for four proteins is displayed in  Fig.~\ref{fig:flatness}(c--f) (see also Fig.~S7) and compared with the random decoys of each of them, matching the different curves at infinite temperature (cf. the insets). In most cases, proteins display in the native energy region (below the transition state) an entropy that is larger than that of the random decoys. This effect is summarized Fig.~\ref{fig:flatness}(a), that displays the density of local entropy
\begin{equation}
    s_{\mathrm{loc}} = \frac{1}{N} \log \sum_{E=E_{\mathrm{N}}}^{E_\ddagger} e^{-\beta E + S(E)},
    \label{eq:le_density}
\end{equation}
where $E_{\mathrm{N}}$ is the minimum energy of the system and $E_\ddagger$ is the energy of the transition state, approximated as the average energy at the transition temperature (cf. caption of Table~S3). The local entropy is calculated at low temperature ($\beta=10^{-1}$) at which all proteins and decoys are stable (cf. Fig.~S8). Equation (\ref{eq:le_density}) is the microcanonical counterpart of Eq. (\ref{eq:locent}). The density of local entropy of native proteins is always larger than the average $\overline{s}$ of the decoys, in five cases out of eight for more than one standard deviation $\sigma_s$. From $\overline{s}$ and $\sigma_s$ we estimated the p--values associated with the null hypothesis that the entropy density of the native protein is smaller than that of the decoys within a Gaussian approximation, that is $p=[1-\text{erf}((s-\overline{s})/\sqrt{2}\sigma_s)]/2$, where $\text{erf}$ is the error function. The p--values are rather low, except for CI2 (2ci2) and for the CHE--Y (1cye).

The protein HHH, designed {\it de novo} and not shaped by natural evolution, seems to display the same features of natural proteins. However, one has to consider that the scaffold they used for the design is a helix bundle typical of natural proteins.

Finally (cf. Fig.~S8, in the Supp. Mat.), we observed that the peaks in the specific heat of the four proteins tend to be wider and less pronounced, in agreement with what we found in subsection \ref{subsect:time}: the wider the specific heat, the less cooperative the transition \cite{Privalov1974} (a measure of cooperativity of the transition can be found in Table~S2 in the Supp. Mat). At the same time, the majority of native proteins display a larger folding temperature (cf. the main peak in the specific heat in Fig.~S8), in agreement with the fact that a larger local entropy stabilizes the native state.

\section{The local entropy depends on the topology of contacts}

Finally, one can investigate whether there is an elementary way to reinterpret why natural proteins display a larger local entropy than random conformations. This aspect can be studied  easily in the context of a Go model in which the frustration associated with the sequence is minimized \cite{Bryngelson1987}.

In fact, in this case one can insert in the definition of local entropy of Eq. (\ref{eq:locent}) the expressions of Eqs. (\ref{eq:ulattice}), (\ref{eq:dist}) and (\ref{eq:j}), obtaining
\begin{align}
    S_{loc}(\Gamma_0)&= \\
    =\log \int & d\Gamma \exp\left[  \left(\beta J+\frac{\gamma}{N_c}\right)\sum_{i<j}^N\Delta_{ij}(\Gamma_0)\Delta_{ij}(\Gamma) \right], \nonumber
\end{align}
where $d\Gamma\equiv dx_1 dx_2\dots$ and an immaterial constant has been disregarded. This is essentially the free energy of the Go model at a rescaled temperature. The exponential in the integrand can be expanded in series,
\begin{align}
    &e^{\dots}=1+\left(\beta J+\frac{\gamma}{N_c}\right)\sum_{i<j}^N\Delta_{ij}(\Gamma_0)\Delta_{ij}(\Gamma)+     \label{eq:exp}\\
    &+\frac{1}{2}\left(\beta J+\frac{\gamma}{N_c}\right)^2\sum_{\substack{i<j\\k<l}}^N\Delta_{ij}(\Gamma_0)\Delta_{kl}(\Gamma_0)\Delta_{ij}(\Gamma)\Delta_{kl}(\Gamma)+\dots \nonumber
\end{align}
Moving the $\Delta_{ij}(\Gamma_0)$ out of the integral and defining the interaction volume as $v\equiv\int d\Gamma \Delta_{ij}(\Gamma)$ and the total volume available to each degree of freedom as $V$, one obtains
\begin{equation}
    S_{\mathrm{loc}}(\Gamma_0) = \log \left[ V^N+ \clusterA+ \clusterB+\clusterA\;\clusterA+\clusterC+\dots \right],
    \label{eq:cluster}
\end{equation}
where the graphs indicate the terms of the expansions. For example,
\begin{equation}
    \clusterA = \left(\beta J+\frac{\gamma}{N_c}\right) V^{N-1}v\sum_{i<j}^N\Delta_{ij}(\Gamma_0)
\end{equation}
is the contribution of a single contact, that depends on the number $\sum\Delta_{ij}(\Gamma_0)$ of contacts of the native conformation. Thus, the larger is the number of contacts on the native conformation, the larger is the local entropy. Similarly,
\begin{equation}
    \clusterB = \frac{1}{2}\left(\beta J+\frac{\gamma}{N_c}\right)^2 V^{N-2}v^2\sum_{i<j<k}^N\Delta_{ij}(\Gamma_0)\Delta_{jk}(\Gamma_0),
\end{equation}
where $\sum\Delta_{ij}(\Gamma_0)\Delta_{jk}(\Gamma_0)$ is the number of triples of nodes (e.g., amino acids) interacting pairwise. For a graph with loops, for example,
\begin{equation}
    \clusterD = \frac{1}{3!}\left(\beta J+\frac{\gamma}{N_c}\right)^3 V^{N-2}A_{\triangleright}v^2\sum_{i<j<k}^N\Delta_{ij}(\Gamma_0)\Delta_{jk}(\Gamma_0)\Delta_{ki}(\Gamma_0),
\end{equation}
where $A_{\triangleright}=3/4$ is a parameter that arise in the integration of the contact functions $\Delta$ from the constraints given by looped graphs. In fact,
\begin{equation}
    \int dx_i dx_j dx_k \Delta(|x_i-x_j|)\Delta(|x_j-x_k|)\Delta(|x_k-x_i|)=V\cdot\frac{3}{4}v^2.
\end{equation}
For a generic graph, $A\leq 1$ and is equal to the unity if the graph does not contain loops because the variables can be integrated sequentially. For a fully--connected graph of $n$ nodes, that in the language of network theory is called a clique, $A=n/2^{n-1}$.

By the linked cluster theorem, the sum of graphs in Eq. (\ref{eq:exp}) is the exponential of the sum of connected graphs, so the local entropy results simply the sum of connected graphs. Note that the connected graphs are different from zero only if the associated $\sum \Delta_{ij}(\Gamma_0)\Delta_{jk}(\Gamma_0)\dots$ is different from zero, that is the corresponding structure is in the native protein. The goal is to spot what are the most important graphs.
The general form of a graph is
\begin{equation}
     \frac{A}{l!}\left(\beta J+\frac{\gamma}{N_c}\right)^l V^{N-n+1} v^{n-1} \cdot (\text{\# of instances in $\Gamma_0$} ),
    \label{eq:graph}
\end{equation}
where $n$ is the number of interacting nodes, $l$ is the number of links (i.e., interactions between nodes) and $A$ depends on how links loop together. Each term is also proportional to the number of instances that the specific graph appears in the native conformation of the protein.

Proteins will display a large local entropy if they are rich of graphs with large values of
\begin{equation}
    w\equiv\frac{A}{l!}B^l V^{N-n+1} v^{n-1},
\end{equation}
where we defined $B\equiv \beta J+\gamma/N_c$.
Thus, local entropy depends on the topology of native contacts. For unlooped graphs ($A=1$), for fixed $n$, $w$ has a maximum at $l^*=B$. Typically, for proteins $\beta J$ is of the order of 1; $\gamma/N_c$ for the model of Fig. \ref{fig:folding_time}b is of the order of $10^4/10^2=10^2$, and consequently $l^*\sim 10^2$.

However, for graphs composed of $l^*$ links, fully--connected graphs display a larger weight $w$. In fact, if we compare the value of $w_{\text{clique}}$ for a fully connected graph ($l\sim n^2$) with that $w_{\text{unloop}}$ of an unlooped graph ($l\sim n$) at $l=B$, one obtains $w_{\text{clique}}/w_{\text{unloop}}=(V^N/v)^{B/2}B^{-1/2}$ that is large for $B\gg 1$.

It was shown \cite{England2003} that $\tr [\Delta_{ij}(\Gamma_0)]^n$ for $n\gg 1$ is a good determinant of protein designability. Since this quantity counts the number of closed paths of length $n$ in the interaction network of the protein, it will be correlated with the number of clusters with large $w$; consequently, the tendency of evolution to maximize $\tr [\Delta_{ij}(\Gamma_0)]^8$ \cite{Tiana2004d} is tantamount to the optimization of the local entropy.
Notice that for the lattice--model proteins discussed above, the average values of $\tr [\Delta_{ij}(\Gamma_0)]^n$ with $n=4,8$ are increasing with $\gamma$, that is with the bias towards having high entropy at short distances  (cf. Table~S4 in the Supp. Mat.).

\section{Discussion and Conclusions}

The concept of local entropy was used to explain the properties of complex systems like artificial neural networks \cite{baldassi2015subdominant,unreasoanable}, in which the "energy" landscape is highly roughed and the system learning dynamics leads to atypical configurations with respect to the Gibbs measure.
In particular, states characterized by a large local entropy can be easily accessed in spite of the abundance of competing states that tend to block the learning procedure.

In the present work we claimed that the concept of local entropy plays an important role also in systems that can move in a smoother energy landscape and thus are not in a glassy regime. In particular, we focused our attention on the three--dimensional conformation of proteins, that evolved along the eons through a complex dynamics described by Darwinian evolution. The high diversity of native conformations of known proteins and their redundancy (i.e., the existence of analogous proteins) suggests that evolution explored a large part of conformational space and thus that protein conformations can be regarded as stationary realizations of the evolutive dynamics.

In the present work, we made use of a Go model to estimate the local entropy of proteins, in order to separate the evolutionary problem of sequence design from that of conformational selection, removing in this way frustration from the system \cite{Das10141}. Although it is known \cite{Paci2002} that frustration plays an important role in determining the features of the denatured state up to the formation of the transition state, one expects it to be less relevant to determine the local properties of the native state and thus the estimation of its local entropy, in agreement with the principle of minimal frustration of the native state \cite{Bryngelson1987}. We thus think that the Go model is particularly suitable for the specific problem we face.

A relevant question is then whether the ensemble of native conformations can be described by Boltzmann statistics or the evolutionary accessibility of native conformations is important to define the fitness of proteins. The latter case would imply that out-of-equilibrium effects affect the set of existing protein conformations.
In the cases we studied, the local entropy of proteins is indeed larger than that of random decoys displaying the same length and density.
As already mentioned in the introduction, these ideas are not new. However, our aim was to provide a simple proof of concept with a novel technique that can be generalized to more realistic settings for protein design. While previous estimation of the local entropy are either qualitative  \cite{Shortle1998a} or based on a knowledge--based, empirical function \cite{Sankar2017}, the methods we suggest inspired by the study of complex systems \cite{baldassi2015subdominant} make use of a direct calculation of the local entropy. Being based on simple statistical--mechanical concepts, we expect this method to be widely applicable to different class of models.

In fact, we  showed that the large local entropy of model proteins has consequences both on their thermodynamic and equilibrium properties. It straightforwardly stabilizes the native state, decreasing its free energy, but extending to the transition state can also improve the folding rate. From this point of view, the concept of evolutionary optimization of the local entropy is related to the dynamical variational principle stated in ref. \cite{Maritan2000}. In fact, also the mechanistic reason for the increase in the folding rate is the same: local entropy biases the stabilizing contacts to be closer along the sequences, making their formation kinetically faster \cite{Plaxco1998}.

Although this point is not within the scope of the present work, we note that the results we discuss could be used in connection with the recent advancements in the understanding of the relation between protein sequence and structure \cite{baek2021accurate,jumper2021highly} to improve the design of new proteins.

\paragraph*{Code availability} The scripts and the main data used and produced in this work can bee freely downloaded at\\ https://gitlab.com/bocconi-artlab/local-entropy-proteins.




\bibliographystyle{unsrt}
\bibliography{main}

\end{document}